\RequirePackage[hyphens]{url}
\documentclass[%
 aip,
 jcp,%
 amsmath,amssymb,
 bibnotes,
 floatfix,
 reprint,%
]{revtex4-1}

\usepackage{hyperref,url}
\usepackage{graphicx}
\usepackage[capitalise]{cleveref}
\usepackage{dcolumn}
\usepackage{bm}
\usepackage{dcolumn}
\hypersetup{hyperindex,breaklinks,colorlinks=true,linkcolor=blue,citecolor=blue,urlcolor=blue,filecolor=blue}

\newcommand{\bk}{\mathbf{k}}

\newcommand{\br}{\mathbf{r}}
\newcommand{\bx}{\mathbf{x}}

\newcommand{\bR}{\mathbf{R}}

\newcommand{\hh}{\hat{H}}

\newcommand{\hc}{\hat{c}}
\newcommand{\hcd}{\hat{c}^{\dagger}}

\begin{document}

\title{Auxiliary-field quantum Monte Carlo calculations of the structural properties of nickel oxide}
\author{Shuai Zhang}
\author{Fionn D. Malone}
\author{Miguel A. Morales}
\email{moralessilva2@llnl.gov}
\affiliation{Lawrence Livermore National Laboratory, Livermore, California 94550, USA}


\begin{abstract}
Auxiliary-field quantum Monte Carlo (AFQMC) has repeatedly demonstrated itself as one of the most accurate quantum many-body methods, capable of simulating both real and model systems.
In this article we investigate the application of AFQMC to realistic strongly correlated materials in periodic Gaussian basis sets.
Using nickel oxide (NiO) as an example, we investigate the importance of finite size effects and basis set errors on the structural properties of the correlated solid. 
We provide benchmark calculations for NiO and compare our results to both experiment measurements and existing theoretical methods. (LLNL-JRNL-752156)
\end{abstract}

\maketitle

\section{\label{sec:intr}Introduction}
Understanding and predicting the properties of strongly correlated materials is one of the grand challenges of modern electronic structure theory.
Such materials exhibit a wealth of exotic phenomena, including magnetism~\cite{magnetism}, metal insulator transitions~\cite{mit1998}, heavy fermion physics~\cite{heavyfermion1,heavyfermion2} and high $T_c$ superconductivity~\cite{highTc1,highTc2}.
Historically, low energy effective theories were developed in an effort to simplify the understanding of these phenomena\citep{Hubbard238}.
However, with the advent of modern supercomputers, there has been a renewed effort to instead describe these materials directly from first principles.

The first-principles description of strongly correlated materials is complicated due to the strong interactions between localized and itinerant electrons.
Moreover, magnetism and superconductivity are inherently many-body effects which are generally poorly described by mean-field approaches.
For example, results from density functional theory (DFT)~\cite{dft1hk,dft2ks} often depend sensitively on the choice of exchange correlation functional.
Although hybrid functionals~\cite{HSE2006,pbe0a,pbe0b,b3lyp1,b3lyp2} and adaptations for strong correlation~\cite{LDAU1998} often yield better results, they rely on additional unknown parameters in the form of the percentage of exact exchange or value of Hubbard $U$.
Motivated by this, there has been significant progress in the development of beyond-DFT methods, such as the GW and Bethe-Salpeter approach~\cite{gwbse}, as well as the dynamical mean field theory (DMFT)~\cite{dmft1996,dmft,dmft1997,Ren2006,dmft2016}.
However, these approaches are mainly used to compute band gaps and optical properties and also introduce additional uncontrolled approximations\cite{Carter800}.

In recent years, there has been a growth in the interest of applying wave-function-based quantum chemistry methods to problems in solid state physics.
This hierarchy of approaches, beginning with Hartree-Fock theory and ending with full configuration interaction (FCI), offer a systematic route to solving the many-electron Schr{\"o}dinger equation directly.
Unfortunately, they come with a cost which is often prohibitively large.
For example, conventional coupled cluster singles and doubles scales like the sixth power of the system size while FCI scales exponentially.
Given that resolving, for example, magnetic correlations requires large simulation cells, it is unclear how useful these methods will be in overcoming the shortcomings of DFT.

Quantum Monte Carlo (QMC) methods offer another route to directly solving the many-electron Schr{\"o}dinger equation with often much more favorable scaling.
Auxiliary field QMC (AFQMC) is one such QMC method that has shown great promise in the simulation of many-body systems\citep{PhysRevB.55.7464,Zhang2003phaseless}. 
Much like conventional quantum chemistry methods, AFQMC works in a second-quantized orbital-based basis which has a number of advantages.
For example, the evaluation of ground state properties other than the total energy are greatly simplified, including dipole moments, reduced density matrices\citep{doi:10.1021/acs.jctc.7b00730}, excited states\citep{1367-2630-15-9-093017,doi:10.1063/1.4861227,PhysRevB.94.085140} and forces \citep{mottaforces}.
Additionally, electron-core interactions can be treated straightforwardly using either pseudo-potentials \citep{Suewattana2007,ma_auxiliary-field_2017} or frozen cores\citep{ma_auxiliary-field_2017,purwanto_downfolding}, while spin-orbit coupling can also be naturally incorporated.
Unfortunately, like most QMC methods, AFQMC is plagued by the fermion sign problem which has no known solution in general.
In order to overcome this, a constraint\citep{PhysRevB.55.7464, Zhang2003phaseless} is usually applied using a trial wave-function which introduces an uncontrolled approximation in the simulations.
However, recent developments using multi-determinants\citep{doi:10.1063/1.3077920,borda2018non}, generalized Hartree-Fock\citep{PhysRevB.94.085103,chang2017multi}, and self-consistently determined trial wave-functions\citep{PhysRevB.94.235119,Zheng1155} have been found to dramatically improve this bias while only modestly increasing the computational effort.
With these advances, AFQMC has demonstrated itself as one of the most accurate methods for simulating strongly correlated model systems\citep{PhysRevX.5.041041,PhysRevX.7.031059,Zheng1155}. 
However, its performance for more realistic strongly correlated materials is less well understood and so far the applications have been limited to a handful of calculations, including the cold curve of copper~\cite{ma_auxiliary-field_2017}
and the spin gap of NiO~\cite{MaPRL2015}.
Here, we apply the phaseless AFQMC method to study the static properties of nickel oxide (NiO), an archetypical, strongly correlated, transition metal oxide.

We note that an alternative approach to AFQMC is the diffusion Monte Carlo\citep{RevModPhys.73.33} (DMC) method.
DMC is formulated in real space which removes the basis set overhead from which AFQMC suffers.
This allows typically larger simulations to be tackled more straightforwardly.
However, DMC also suffers from a number of issues associated with improving trial wave-functions and the use of non-local pseudo-potentials~\cite{PhysRevB.93.094111}.
Nonetheless, it remains a promising and complementary approach to AFQMC in the study of strongly correlated realistic materials\citep{Schiller2015,krogelDMC,wagner_ceperley,doped_def_nio}.

This paper is organized as follows. In  \cref{sec:meth} we outline the basics of the phaseless AFQMC algorithm and discuss some specific implementation details relevant to efficiently applying it using periodic Gaussian basis sets.
In \cref{sec:result} we present benchmark AFQMC results for a four-atom cell of NiO and investigate finite size and basis set errors. 
Finally, in \cref{sec:conclude}, we discuss the future prospects of AFQMC as a predictive tool for studying strongly correlated materials.

\section{\label{sec:meth}Methodology}
In this section we briefly outline the phaseless AFQMC algorithm\citep{Zhang2003phaseless,motta2017ab}.
Although AFQMC is applicable to a wide variety of real and model systems, here we focus on its application to periodic solids in Gaussian basis sets.
\subsection{Overview of AFQMC}

We are interested in solving for the ground state of a generic many-electron Hamiltonian, which can be written in second-quantized form as
\begin{align}
    \hat{H} &= \sum_{ij\sigma}^M h_{ij} \hcd_{i\sigma}\hc_{j\sigma} + \frac{1}{2}\sum_{ijkl\sigma\sigma'}^M v_{ijkl}  \hcd_{i\sigma}\hcd_{j\sigma'}\hc_{l\sigma'}\hc_{k\sigma}+E_{II},\label{eq:hamil}\\
            &= \hh_1 + \hh_2 + E_{II}
\end{align}
where $M$ is the number of single-particle states in our basis, $E_{II}$ is the energetic contribution from the static ionic configuration, and $\hcd_{i\sigma}$ and $\hc_{i\sigma}$ create and annihilate an electron in some single-particle state $|i\sigma\rangle$, where $\sigma$ is the electron's spin. The one- and two-body matrix elements are given by
\begin{equation}
h_{ij} = \int d\br \  \varphi_{i}^*(\br)\left(-\frac{1}{2}\hat{\nabla}_\br^2 - \sum_{I}\frac{Z_I}{|\br-\bR_I|}\right)\varphi_{j}(\br),\label{eq:hij}
\end{equation}
where $\langle \br | i\rangle = \varphi_{i}(\br)$, $Z_I$ and $\bR_I$ are the ionic charge and position of the atom $I$ respectively, and
\begin{equation}
    v_{ijkl} = \int \int d\br \ d\br' \ \varphi^*_i(\br)\varphi^*_j(\br')\frac{1}{|\br-\br'|}\varphi_{k}(\br)\varphi_l(\br')\label{eq:four_ix},
\end{equation}
are the two-electron repulsion integrals. For calculations with core electrons, the electron-ion Coulomb interaction in \cref{eq:hij} can be replaced by any desired approximation (e.g. pseudo-potential, effective core potential, frozen core, etc)\citep{Suewattana2007,ma_auxiliary-field_2017,ma_auxiliary-field_2017,purwanto_downfolding}.
Hartree atomic units are used throughout.

One way to find the ground state, $|\Psi_0\rangle$, of $\hh$ is to use the projection method:
\begin{equation}
|\Psi_0\rangle \propto \lim_{\tau\rightarrow\infty} e^{-\tau\hat{H}}|\phi\rangle,\label{eq:projection}
\end{equation}
where $|\phi\rangle$ is some initial state (here a Slater determinant) satisfying $\langle \phi | \Psi_0\rangle \ne 0$.
In practice, the long time limit of \cref{eq:projection} can be found iteratively using
\begin{equation}
    |\Psi^{(n+1)}\rangle = e^{-\Delta\tau\hh} |\Psi^{(n)}\rangle\label{eq:iterate},
\end{equation}
where $\Delta\tau$ is the time step.
In order to proceed, we need to find an efficient way to apply the imaginary time propagator in \cref{eq:iterate}.
There are many different ways to achieve this, each generally leading to a different QMC algorithm\citep{RevModPhys.73.33,fciqmc2009}.
We first split up the one- and two-body Hamiltonian in the exponential in \cref{eq:iterate} and use the second-order Suzuki-Trotter decomposition
\begin{equation}
e^{-\Delta\tau\hat{H}} = e^{-\frac{\Delta\tau}{2} \hat{H}_1}e^{-\Delta\tau \hat{H}_2}e^{-\frac{\Delta\tau}{2} \hat{H}_1} + \mathcal{O}(\Delta\tau^2)\label{eq:suz_trot}.
\end{equation}
In AFQMC we represent the many-electron wave-function in a basis of non-orthogonal Slater determinants.
The action of the exponential of a one-body operator on a Slater determinant yields yet another Slater determinant by Thouless' theorem\citep{thouless_stability_1960}.
However, no simple relationship exists in general for the exponential of a two-body operator.
To overcome this, we can write the two-body Hamiltonian in \cref{eq:hamil} as
\begin{equation}
    \hat{H}_2 = -\frac{1}{2} \sum_{\gamma} \hat{v}_{\gamma}^2 + \hat{v}_0\label{eq:hsq},
\end{equation}
where $\hat{v}_{\gamma}$ is a one-body operator, and use the Hubbard-Stratonovich transformation\cite{PhysRevLett.3.77} to write
\begin{equation}
e^{\frac{\Delta\tau}{2}\sum_{\gamma}\hat{v}_{\gamma}^2} = \prod_\gamma \int d x_\gamma e^{-\frac{x_\gamma^2}{2}} e^{\sqrt{\Delta\tau} x_\gamma\hat{v}_\gamma} \label{eq:hs}.
\end{equation}
Inserting \cref{eq:hs} into \cref{eq:suz_trot}, we have
\begin{equation}
    |\Psi^{(n+1)}\rangle =   \int d \bx p(\bx)\hat{B}(\bx) |\Psi^{(n)}\rangle, \label{eq:iterate_hs}
\end{equation}
where $\hat{B}(\bx)$ now contains exponentials of one-body operators only.
The multi-dimensional integral in \cref{eq:iterate_hs} can be evaluated using Monte Carlo integration over normally distributed auxiliary fields $\bx$.
In practice, we instead express our wave-function as a sum over weighted random walkers
\begin{equation}
    |\Psi^{(n)}\rangle = \sum_\alpha^{N_w} w^{(n)}_\alpha |\phi^{(n)}_\alpha\rangle,
\end{equation}
where $w^{(n)}_\alpha$ is the walker's weight at time step $n$ and $|\phi^{(n)}_\alpha\rangle$ is the walker's Slater determinant.
Solving \cref{eq:iterate} then amounts to repeatedly propagating the walker's Slater determinant by $\hat{B}(\bx)$ and updating the walker's weights appropriately.

Unfortunately, this ``free-projection'' algorithm suffers from a serious phase problem.
In the long imaginary time limit of the propagation, one finds that the walker's weights are uniformly distributed in the complex plane, thus rendering the accumulation of statistics essentially impossible.
This is a manifestation of the notorious fermion sign problem which has no known solution in general.
To overcome this, Zhang \emph{et al.}\cite{Zhang2003phaseless} introduced the phaseless approximation to control the walker's phase and render the walker's weights positive, leading to a stable numerical algorithm at the cost of introducing a systematic bias.

In the phaseless AFQMC approach we rewrite the propagation as
\begin{equation}
    |\Psi^{(n+1)}\rangle =   \int d \bx p(\bx) I(\bx,\bar{\bx},|\phi\rangle) \hat{B}(\bx-\bar{\bx})|\Psi^{(n)}\rangle, \label{eq:iterate_imp}
\end{equation}
where
\begin{equation}
I(\bx,\bar{\bx},|\phi\rangle) = \frac{\langle \psi_T|\hat{B}(\bx- \bar{\bx}) |\phi\rangle}{\langle \psi_T|\phi\rangle} e^{\bx\cdot\bar{\bx}-\frac{\bar{\bx}\cdot\bar{\bx}}{2}}
\end{equation}
is the importance function, $\bar{\bx}$ is the ``force-bias'' shift and $|\psi_T\rangle$ is a trial wave-function.
The optimal force-bias term, which cancels fluctuations in the importance function to $\mathcal{O}(\sqrt{\Delta\tau})$, can be shown to be\cite{Zhang2003phaseless}
\begin{equation}
    \bar{x}_\gamma = -\sqrt{\Delta\tau}\frac{\langle \Psi_T| \hat{v}_\gamma | \phi \rangle}{\langle \Psi_T | \phi \rangle}.
\end{equation}
At this point \cref{eq:iterate_imp} is still exact.
The importance function encourages walkers to areas of the Hilbert space with a larger overlap with the trial wave-function.
However, the reformulation is only useful in order to eventually impose a constraint.
As before, a given walker's Slater determinants is propagated by \mbox{$\hat{B}(\bx-\bar{\bx})$}, but now its weight is modified:
\begin{equation}
    w_\alpha^{(n+1)} = |I(\bx,\bar{\bx},|\phi^{(n)}_\alpha\rangle)|\times \max \left(0, \cos \Delta \theta\right) w_\alpha^{(n)},
\end{equation}
where the phase is defined as
\begin{equation}
    \Delta \theta = \arg\left(\frac{\langle \psi_T|\hat{B}(\bx- \bar{\bx}) |\phi_\alpha^{(n)}\rangle}{\langle \psi_T|\phi_\alpha^{(n)}\rangle}\right).
\end{equation}
Thus, the walker's weights remain positive and those walkers with rapidly changing phases are killed and removed from the simulation.
The trial wave-function now takes a central position in the algorithm by imposing the constraint.
The constraint can be systematically improved by using better trial wave-functions but often a single Slater determinant of Hartree-Fock or DFT orbitals is found to yield highly accurate energies.

\subsection{Implementation Details \label{sec:implement}}
The above formulation of AFQMC has been applied to a wide variety of problems in quantum chemistry and solid state physics\citep{doi:10.1021/acs.jctc.7b00730,Alsaidi2006TMOmol,Alsaidi2006Gaussianbasis,Alsaidi2006postd,Alsaidi2006H,doi:10.1063/1.3077920,Purwanto2009Si,1367-2630-15-9-093017,mottaforces}.
Previous application of AFQMC in solids have mainly employed plane wave basis sets which have the primary advantage of simplifying both the Hubbard-Stratonovich transformation and the evaluation of matrix elements of the Hamiltonian\citep{Suewattana2007}.
Additionally, plane waves and pseudo-potentials form the bedrock of most electronic structure methods, so decades of experience can be built upon\citep{ma_auxiliary-field_2017}.
Despite these advantages, often prohibitively large plane wave expansions are required to converge the total energy.
Therefore, we seek a more compact basis set which can better represent the localized $d$ and $f$ orbitals which play such an important role in the physics of strong correlation.
Note that the downfolding approach of Ref.\citenum{MaPRL2015} can also reduce the size of basis sets required.

Fortunately, there has been a resurgence in interest in the application of wave-function based quantum chemistry methods to solids in recent years\citep{gillan_high-precision_2008,nolan_calculation_2009,booth_towards_2013,mcclain_gaussian-based_2017,doi:10.1063/1.4998644}.
This, in turn, has lead to the development of robust periodic Gaussian basis sets which we adapt for use in AFQMC in this work.
Explicitly, we use a basis of periodic atomic orbitals
\begin{equation}
    \varphi_{n\bk}(\br) = \sum_{\mathbf{T}} e^{i\bk\cdot{\mathbf{T}}}\chi_n(\br-\mathbf{T}),
\end{equation}
where $\chi_n(\br)$ is an atomic orbital, $\bk$ is the crystal momentum and the sum is over translation vectors $\mathbf{T}$ up to a cutoff.
We use the PySCF quantum chemistry package\citep{PYSCF2017} to compute the one- and two-electron integrals and the trial wave-function.
To avoid the $\mathcal{O}(M^4)$ cost of storing $v_{ijkl}$ we use the modified Cholesky decomposition~\citep{beebe77,doi:10.1063/1.1578621,doi:10.1002/jcc.21318,doi:10.1063/1.3654002} to write
\begin{equation}
    V_{(ik),(lj)} = v_{ijkl} \approx \sum_\gamma^{N_\gamma} L_{ik}^\gamma L_{lj}^{*\gamma},\label{eq:cholesky}
\end{equation}
where the number of Cholesky vectors $N_\gamma = c_\gamma M$ is an additional convergence parameter. Typically we find that $c_\gamma\approx 10$ is sufficient for an maximum error of $10^{-5}$ Ha in the integrals.
A similar value of $c_\gamma$ is found for the case of molecular calculations\citep{doi:10.1063/1.3654002,doi:10.1021/acs.jctc.7b00730} where the two-electron repulsion integrals are real.
Note the order of the $jl$ indices are flipped in \cref{eq:cholesky} which is required to ensure that the matrix $V$ is Hermitian and can be Cholesky decomposed.
To perform the Hubbard-Stratonovich transformation we define the Hermitian operators
\begin{align}
    \hat{v}_{\gamma+} &= \sum_{ik\sigma} \left(\frac{L^{\gamma}_{ik}+L^{\gamma*}_{ki}}{2}\right)\hcd_{i\sigma}\hc_{k\sigma} \\
    &=  \sum_{ik\sigma}\left[L_{+}\right]_{ik}^{\gamma}\hcd_{i\sigma}\hc_{k\sigma} \\
\hat{v}_{\gamma-} &= i\sum_{ik\sigma} \left(\frac{L^{\gamma}_{ik}-L^{\gamma*}_{ki}}{2}\right)\hcd_{i\sigma}\hc_{k\sigma}\\
&=  \sum_{ik\sigma}\left[L_{-}\right]_{ik}^{\gamma}\hcd_{i\sigma}\hc_{k\sigma},
\end{align}
so that we can write
\begin{align}
    \hh_2 &= \frac{1}{2} \sum_\gamma \left(\hat{v}^{2}_{\gamma+} +\hat{v}^2_{\gamma-}\right) + \hat{v}_0,
\end{align}
which will lead to $2c_\gamma M$ auxiliary fields.

The force bias term can now be evaluated as
\begin{equation}
    \bar{x}^{\alpha}_{\gamma\pm} = -\sqrt{\Delta\tau}\sum_{ik\sigma}\left[L_{\pm}\right]^{\gamma}_{ik}G^{\alpha}_{i\sigma k\sigma},
\end{equation}
where the walker's Green's function is
\begin{align}
    G^{\alpha}_{i\sigma j\sigma'} &= \frac{\langle\psi_T|\hcd_{i\sigma}\hc_{j\sigma'}|\phi_\alpha\rangle}{\langle\psi_T|\phi_\alpha\rangle}\\
                           &= \left[U_{\sigma'}(V_\sigma ^{\dagger}U_{\sigma'})^{-1}V_{\sigma}^{\dagger}\right]_{ji}\\
    &=\left[V_\sigma^*(U_{\sigma'}^T V_\sigma^*)^{-1}U^{T}_{\sigma'}\right]_{ij}\label{eq:gf},
\end{align}
and $U_\sigma$ and $V_\sigma$ are the Slater matrices of the walker and the trial wave-function respectively.
The cost of evaluating the force-bias potential can be reduced by precomputing some tensors\citep{motta2017ab}.
If we write the Green's function in \cref{eq:gf} as
\begin{align}
    G_{i\sigma j\sigma'}^{\alpha} &= \left[V^{*}_\sigma \mathcal{G_{\sigma\sigma'}} \right]_{ij}
\end{align}
and define the partially contracted Cholesky vector
\begin{equation}
    \left[\mathcal{L}_{\pm}\right]^{\gamma}_{ak\sigma} = \sum_i \left[V^{*}_\sigma\right]_{ia}\left[L_{\pm}\right]^{\gamma}_{ik},
\end{equation}
then we can write\cite{motta2017ab}
\begin{equation}
    \bar{x}_{\gamma\pm} = -\sqrt{\Delta\tau}\sum_{ak\sigma}\left[\mathcal{L}_{\pm}\right]^{\gamma}_{ak\sigma}\mathcal{G}_{a\sigma k\sigma}.
\end{equation}
This brings the cost of computing the force-bias down from $\mathcal{O}(N_\gamma M^2)$ to $\mathcal{O}(N_\gamma NM)$ since $\mathcal{L}_{\pm}^{\gamma}$ can be computed once at the start of the simulation at the cost of $\mathcal{O}(N_\gamma M)$ operations.

Once the system has equilibrated we will have a statistical representation of the approximate ground state wave-function
\begin{equation}
    |\Psi^{n}_0\rangle = \sum_\alpha w^n_\alpha \frac{|\phi^n_\alpha\rangle}{\langle \Psi_T|\phi^n_\alpha\rangle},
\end{equation}
from which we can compute estimates of observables.
For example, the ground state total energy can be computed from the mixed estimator
\begin{align}
    E_{\mathrm{mixed}} &= \frac{\langle \psi_T | \hat{H} |\Psi_0\rangle}{\langle \Psi_T| \Psi_0\rangle} \\
                       &= \frac{\sum_\alpha w_\alpha E_L[\phi_\alpha]}{\sum_{\alpha}w_\alpha},
\end{align}
where the local energy is defined as
\begin{equation}
    \begin{split}
        E_L[\phi_\alpha] &= \sum_{ij\sigma} h_{ij}G^{\alpha}_{i\sigma j\sigma} + \\
        &\sum_{ijkl\gamma\sigma\sigma'}  L_{ik}^\gamma L_{lj}^{*\gamma}\left(G^{\alpha}_{i\sigma
                k\sigma}G^{\alpha}_{j\sigma' l\sigma'}-G^{\alpha}_{i\sigma
                l\sigma'}G^{\alpha}_{j\sigma' k\sigma}\right).
    \end{split}
\end{equation}
To avoid an $\mathcal{O}(M^4)$ evaluation cost of the two-body part of the local energy we again first pre-contract the trial wave-function with the integrals to construct
\begin{equation}
    \begin{split}
    \mathcal{V}_{(ak),(lb)}^{\sigma\sigma'} = \sum_{\gamma} \sum_{ij} L_{ik}^{\gamma} L_{lj}^{\gamma *} \Big(&\left[V^{*}_\sigma\right]_{ia} \left[V^{*}_{\sigma'}\right]_{jb}  - \\
    & \delta_{\sigma\sigma'}\left[V^{*}_\sigma\right]_{ib} \left[V^{*}_{\sigma'}\right]_{ja}\Big).
    \end{split}
\end{equation}
$\mathcal{V}$ requires the storage of at most $2 N^2 M^2$ elements and is constructed once at the start of a simulation.
However, $\mathcal{V}$ is usually a very sparse matrix, so that this storage requirement can be brought down to $\mathcal{O}(s N^2 M^2)$.
Note that for by making use of Bl\"och's theorm, the sparsity is guaranteed to be at least $N_k^{-1}$ where $N_k$ is the number of $k$-points.
We can then calculate the two-body energy as
\begin{equation}
        E_{2B} = \sum_{abkl\sigma\sigma'} \mathcal{V}^{\sigma\sigma'}_{(ak),(bl)}\mathcal{G}^\alpha_{a\sigma k\sigma}\mathcal{G}^{\alpha}_{b\sigma' l\sigma'}
\end{equation}
at the cost of $\mathcal{O}(sN^2M^2)$ operations.
Expectation values of operators which do not commute with the Hamiltonian can be computed using back propagation\cite{PhysRevB.55.7464,PhysRevE.70.056702,doi:10.1021/acs.jctc.7b00730}.

\section{\label{sec:result}Results}
In this section we apply the phaseless AFQMC method to NiO, a prototypical strongly correlated materials.
This system has been of great interest both theoretically~\cite{Tran2006,Feng2004,krogelDMC,Cohen1997,dmft2016,Kobayashi2008,Eder2015} and experimentally~\cite{exptAFMII1,exptAFMII2,exptAFMII3,exptAFMII4,ExptEtoDAC,ExptNOGUCHI1999509,ExptShen1991}.
Under ambient conditions, the type-II anti-ferromagnitic (AFM II) phase of NiO in the rock-salt (B1) structure is found experimentally to be most stable~\cite{exptAFMII1,exptAFMII2,exptAFMII3,exptAFMII4}.
In this phase, each atom is in an octohedral crystal field with Ni having
opposite spins in adjacent atomic planes along the [111] direction.
Previous studies suggest the system to be an
insulator with mixed Mott-Hubbard and 
charge-transfer characteristics~\cite{Tran2006,Schuler2005,OlaldeVelasco2011}.
Theoretical calculations in different
levels (DFT~\cite{Cohen1997}
and DMFT~\cite{dmft2016}) 
uniformly predict a gradual magnetic collapse
and metallization
under large enough compression.
However, the critical compression ratio 
associated with the magnetic 
and metal-insulator transition vary depending on
the specific simulation method used~\cite{Feng2004}.

Here we focus on the insulating phase. We simulate a four-atom cell, the smallest unit cell capable of exhibiting AFM II order, but still challenging to simulate using existing quantum chemistry or many-body methods.
Our goal is to investigate how well AFQMC performs when applied to real strongly-correlated materials, and to investigate the importance of finite size effects and basis set errors.

\subsection{Computational Setup}
We use the PySCF software package\citep{PYSCF2017} to calculate all the input to the AFQMC calculations, including the 1-body hamiltonian, the Cholesky factorized
2-electron integrals and the trial wave-function, which was constructed using the unrestricted Hartree-Fock solution for the AFM II state.
All simulations were performed using Goedecker-Teter-Hutter
(GTH)~\cite{GTH1996} type pseudo-potentials constructed with the
Perdew-Burke-Ernzerhof (PBE)~\cite{PBE1996} exchange-correlation functional, as supplied by the CP2K\citep{doi:10.1063/1.2770708,CP2K2013} software package.
The Ni pseudo-potential treats semi-core states explicitly as the valence electrons, leading to an 18-electron pseudo-potential..
We used the accompanying MOLOPT-GTH DZVP, TZVP, and TZV2P Gaussian basis sets, also from the CP2K distribution.~\footnote{ For Ni, we use are short-range basis sets, MOLOPT-SR-GTH, which are more appropriate for solid state calculations. For O, the short-range basis is available only for DZVP, therefore we use the regular basis set (non-SR ones) for TZVP and TZV2P calculations. Our $\Gamma$-point calculations show that the difference in the cold curve when switching from DZVP-MOLOPT-SR-GTH to DZVP-MOLOPT-GTH basis for O is negligible and leads to changes in $V_0$ and $B_0$ by only 0.4\% and 3 GPa, respectively.}.
All AFQMC calculations were performed using the open-source QMCPACK software package\citep{qmcpack}.
We used $\sim$1000 walkers and a timestep of 0.005 which we found sufficient to control any potential population control and finite timestep biases respectively.
\subsection{Finite Size Effects}

All many-body simulations of finite periodic systems suffer from finite size errors\citep{Drummond2008,PhysRevB.94.035126}.
Typically these are split into one-body and two-body size effects.
One-body errors are related to the underlying single-particle energies and can be removed using twist averaging\citep{twtavr2001}.
Two-body errors have no analogue with mean field theories and contain all size effects which remain after one-body errors have been corrected.
In the past 20 years, numerous approaches have been developed to alleviate these two-body finite size errors\citep{PhysRevB.53.1814,PhysRevLett.97.076404}.
Here we investigate the performance of the corrections developed by Kwee, Zhang and Krakauer\citep{KZK2008} (KZK) and their generalization for magnetic systems\citep{KZK2011}.

The KZK correction is found by computing the difference
between the DFT energy in the infinite supercell size limit
($E_\text{DFT}(\infty)$) and that obtained using
the supercell size-dependent exchange-correlation
functional ($E_\text{DFT}^\text{FS}(L)$). The difference
$\Delta E^\text{DFT}=E_\text{DFT}(\infty)-E_\text{DFT}^\text{FS}(L)$
is applied to the QMC energies to obtain results
which should be closer to the true thermodynamic limit value.
The KZK approach has the advantage that shell effects in the KZK energies at different twist vectors are usually correlated with those in the QMC simulations.
They can therefore be used as a control variate to accelerate the convergence of twist averaging procedure\citep{PhysRevB.88.085121,doi:10.1063/1.4922619}.

\begin{figure}
  \includegraphics[width=0.48\textwidth]{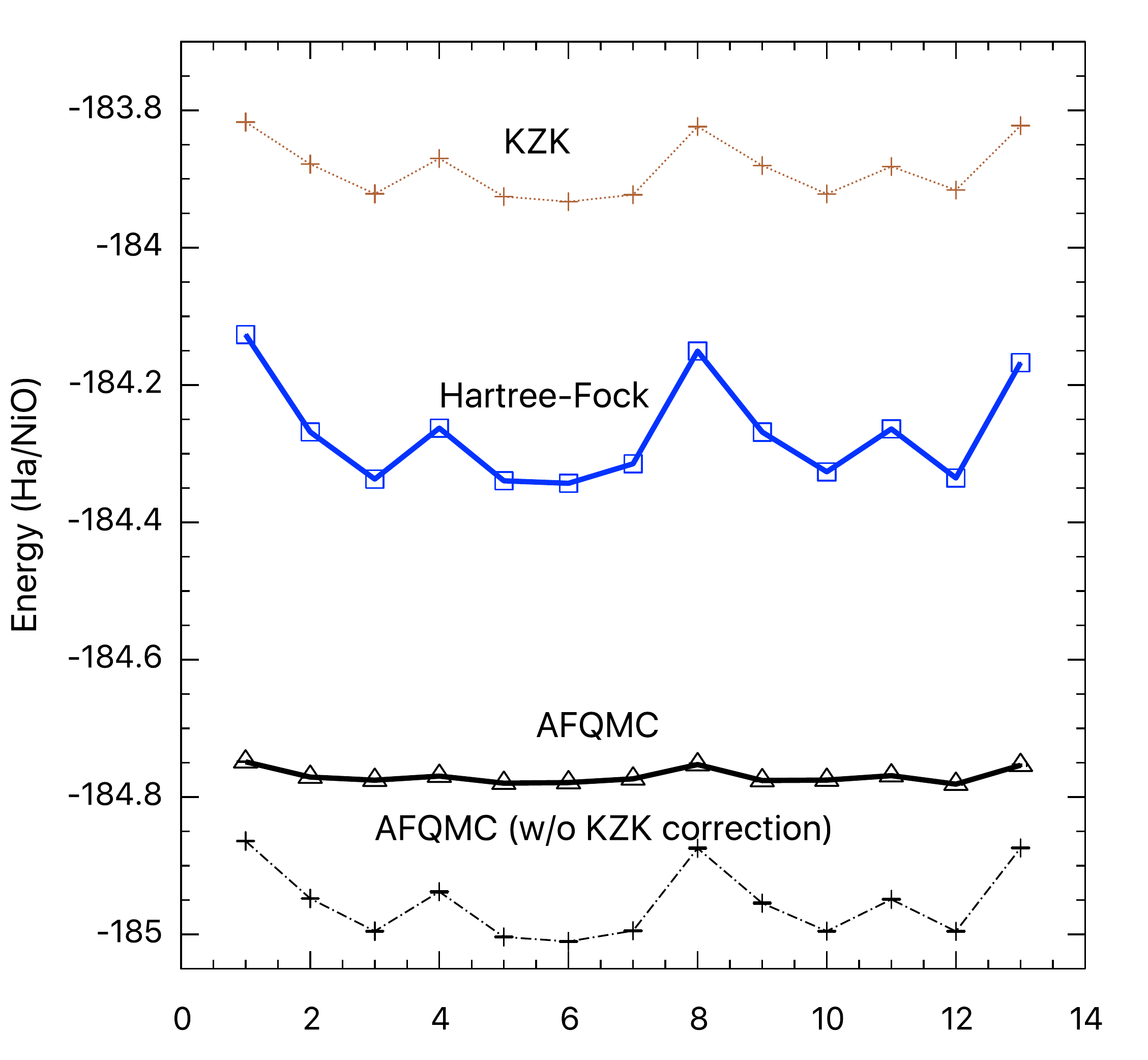}
  \caption{Comparison of the DFT, Hartree-Fock and AFQMC energy as a function of symmetrically inequivalent twist vector index for the four-atom NiO cell at the ambient volume. The twist vectors are chosen from a $\Gamma$-centered $4\times4\times4$ Monkhorst-Pack~\cite{MPk1,MPk2} grid. The DFT simulations were performed using the KZK functional~\cite{QE2009,QE2017,KZK2011} in a plane wave basis set while the Hartree-Fock and AFQMC simulations used the TZV2P basis. The KZK data have been shifted by -125.2 Ha/NiO for clarity.}
  \label{fig:etwist_av}
\end{figure}

In \cref{fig:etwist_av} we compare the AFQMC, KZK and Hartree-Fock energy as a function of the twist vector at the experimental equilibrium lattice constant (4.171 \AA).
We note that while the Hartree-Fock energies exhibits a similar behavior to AFQMC, the KZK energies follow the QMC energies more closely.
Thus, the KZK-corrected AFQMC energy is much smoother allowing for a faster convergence of the twist averaging procedure.
This result suggests that the use of the KZK corrections is justified even in this strongly correlated material, at least when both DFT and AFQMC predict the system to be in the same phase.

\begin{figure}
  \includegraphics[width=0.48\textwidth]{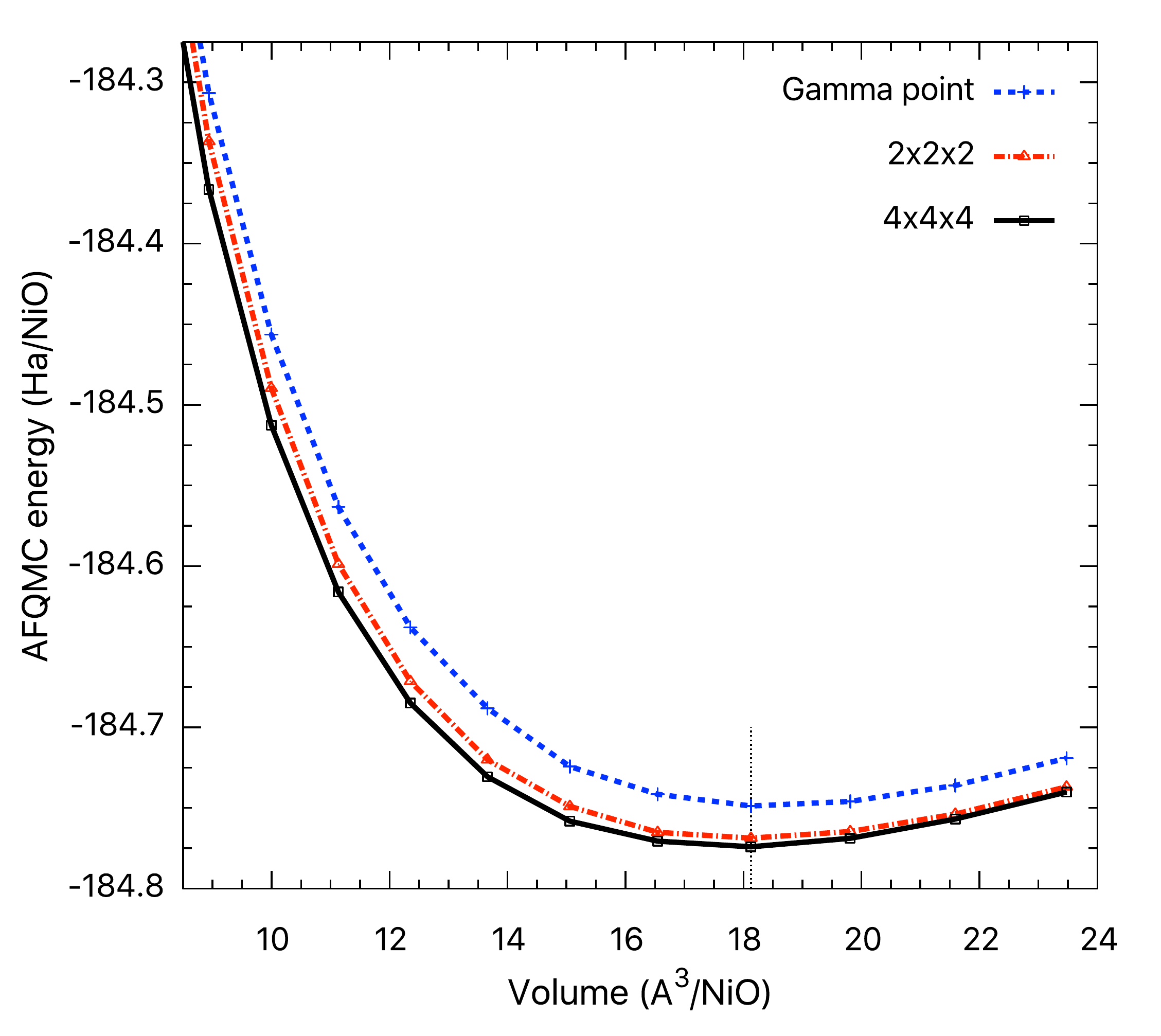}
  \caption{Comparison of the AFQMC cold curve for AFM-NiO obtained using different densities of twist vectors in the TZV2P basis. Curves are guides to the eyes. The dotted vertical line denotes the experimental value for the equilibrium volume.~\cite{Bartel1971}}
  \label{fig:coldc_basis}
\end{figure}

In \cref{fig:coldc_basis} we investigate the convergence of the AFQMC energy with respect to twist averaging as a function of volume.
We see that a finer grid of twist vectors is required at higher densities (lower volumes).
This can be understood as the system becomes more metallic and thus shell effects become more important.

\begin{figure}
  \includegraphics[width=0.48\textwidth]{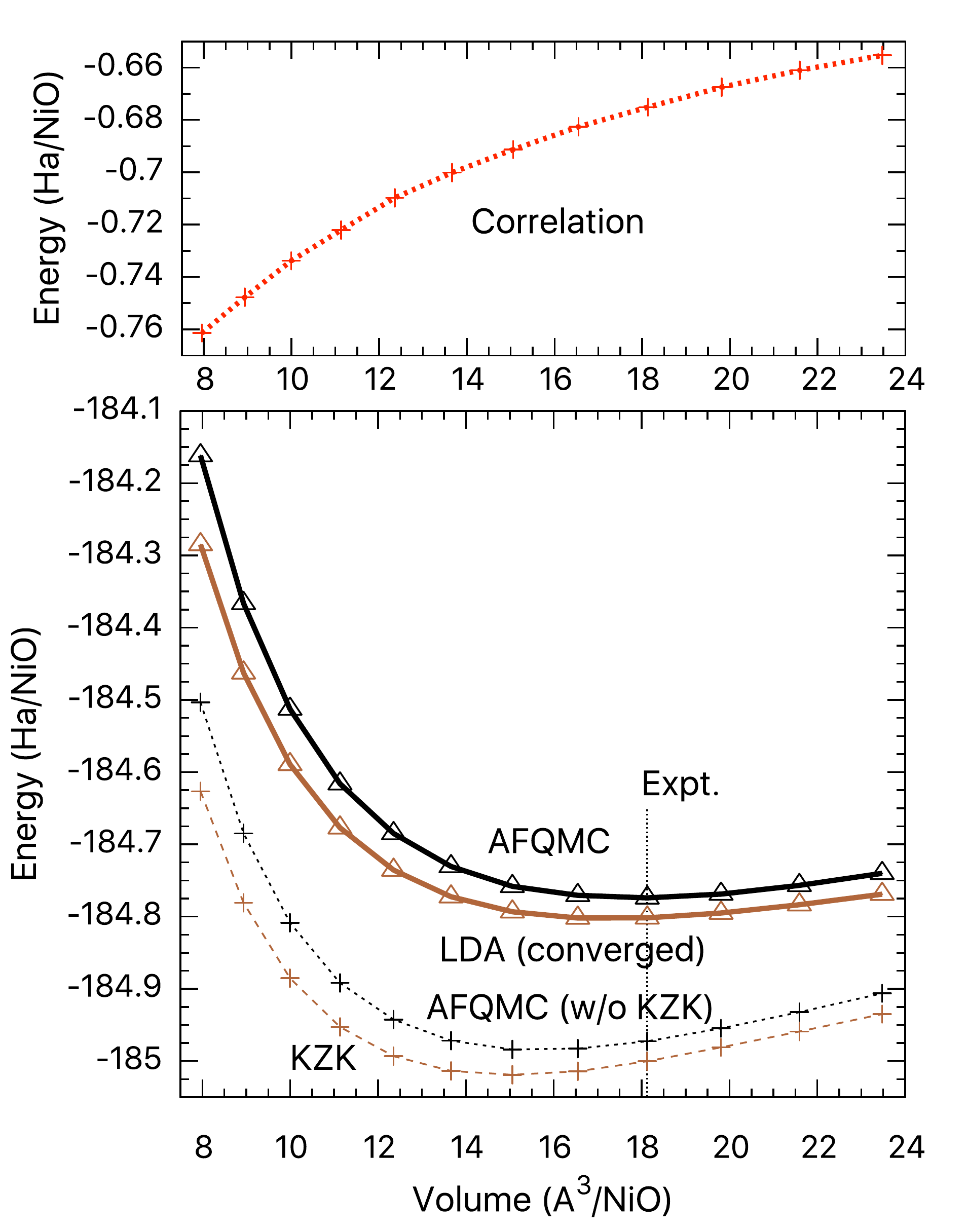}
  \caption{Comparison of cold curves calculated within DFT [using the KZK and local density approximation (LDA) functionals] and AFQMC for the 4-atom cell of NiO in the AFM-II state. The AFQMC simulations were performed using the TZV2P basis\citep{CP2K2013}. The KZK and AFQMC energies have been twist-averaged using a $\Gamma$-centered 4$\times$4$\times$4 $k$ Monkhorst-Pack grid. The converged LDA+U energies were calculated using a $\Gamma$-centered 8$\times$8$\times$8 Monkhorst-Pack grid. The dotted vertical line denotes the experimental value for the equilibrium volume.~\cite{Bartel1971} For clarity, LDA and KZK data have been shifted by -126.3 Ha/NiO. The curves joining the points are meant as guides to the eye.}
  \label{fig:coldc_sc}
\end{figure}

In \cref{fig:coldc_sc} we compare the raw and size-corrected AFQMC and KZK cold curves.
We see that the KZK corrections generally shift the minimum of the AFQMC cold curve towards the experimental volume.
However, the KZK corrections for this small supercell are still quite large.
Larger simulations are required before the accuracy of AFQMC relative to experiment can be safely determined.
Also plotted is the subplot of \cref{fig:coldc_sc} is the correlation energy for the finite supercell.

\subsection{Basis Set Convergence}

\begin{figure}
  \includegraphics[width=0.48\textwidth]{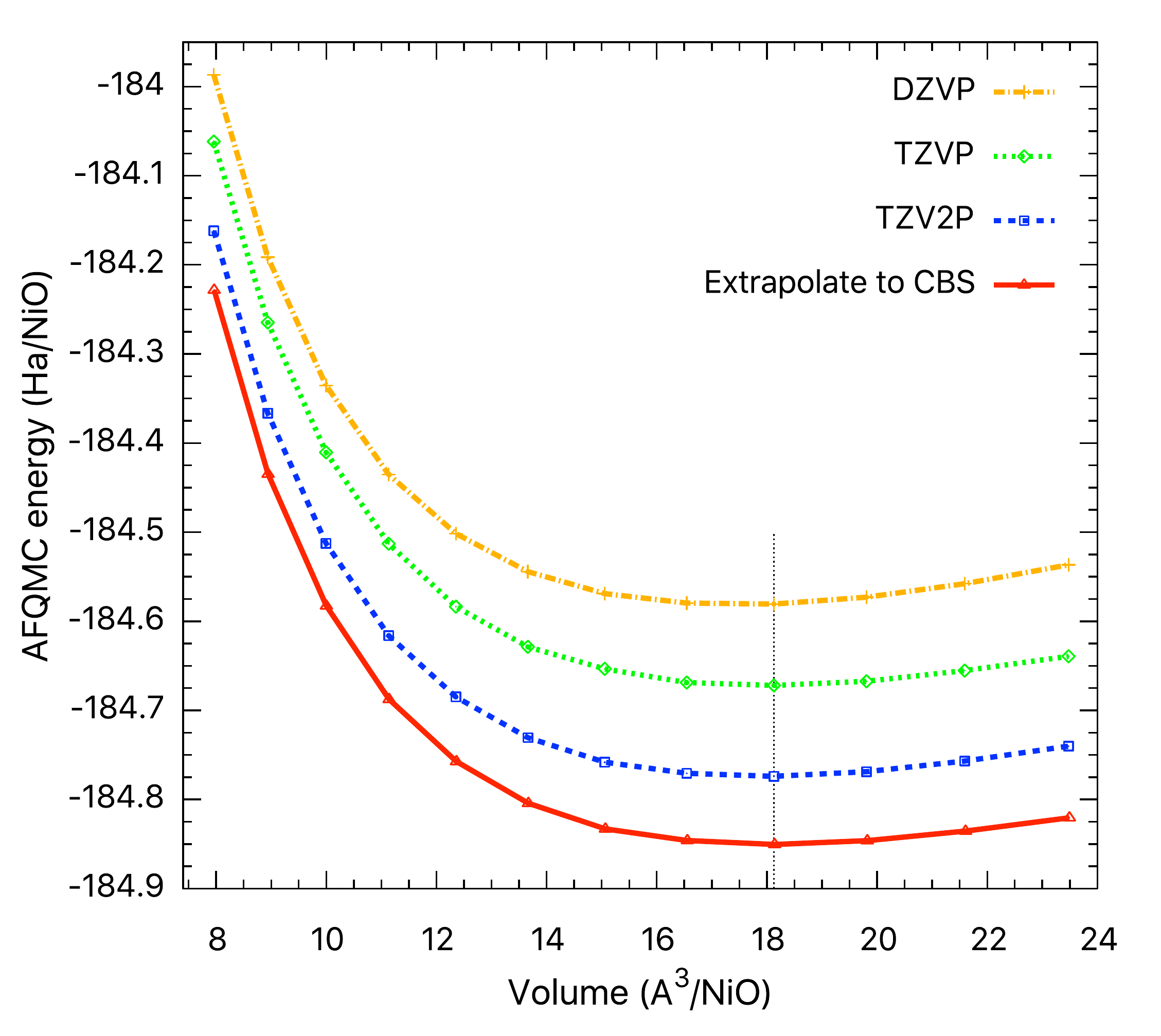}
  \caption{Comparison of AFQMC cold curve for AFM-NiO obtained using different basis sets. Results have been twist-averaged over a $\Gamma$-centered 4$\times$4$\times$4 $k$ grid and size-corrected with the KZK method. Curves are guides to the eyes. The dotted vertical line denotes the experimental value for the equilibrium volume.~\cite{Bartel1971}}
  \label{fig:afqmc_basis}
\end{figure}

We next investigate the dependence of the AFQMC energy on basis set and the corresponding convergence
rate of structural properties.
\cref{fig:afqmc_basis} shows a comparison of the NiO cold curve, as calculated by AFQMC,
for the various basis sets considered in this work; KZK size corrections have been applied and twist averaging was employed
using a 4x4x4 twist grid. As expected, there is a systematic reduction in total energy as the basis set increases in size.
From the figure it is clear that larger basis sets increase the equilibrium volume of the material, bringing results in closer
agreement to experimental measurements. The change in equilibrium volume is large when moving from the DZVP 
to the TZVP basis, with results of 17.49 and 17.92 \AA$^3$/NiO respectively. The change from TZVP to TZV2P is much smaller, 
TZV2P also having a volume of 17.92 \AA$^3$/NiO. While the latter basis set is fairly close to convergence with respect to the complete basis set (CBS)
limit, it is possible to obtain a reasonably accurate estimate of the bulk properties at the CBS limit by employing
a standard basis set extrapolation scheme, very common in the quantum chemistry community when Gaussian basis sets
are employed. In particular, we use the following formula to extrapolate the correlation energy contribution of the energy,
\begin{equation}
    E_c(l_\text{max})=E_c^\text{CBS}+Al_\text{max}^{-3}, \label{eq:cbs}
\end{equation}
where $l_\text{max}$ denotes the highest angular momentum 
included in the basis set.
The AFQMC energies obtained from the extrapolated values of the correlation energy, $E_c^\text{CBS}$, are shown in \cref{fig:afqmc_basis}
with a solid red curve. Several things must be mentioned at this point regarding the extrapolated energies.
First, the TZV2P basis lacks a basis function with angular momentum $l=4$, which is typically included in a triple-zeta quality
basis set in calculations of finite molecular systems. This would somewhat affect the accuracy of the resulting energy extrapolation.
In addition, typical extrapolation schemes in molecular calculations are based on three or more basis sets, in order
to obtain highly accurate extrapolations to the CBS limit. 
Unfortunately, the lack of available basis sets beyond TZV2P prevents us from obtaining more accurate extrapolations at this time.
Nonetheless, given the small magnitude of the correction and the fact that we are mainly interested in the volume dependence only (not in the total magnitude), 
we believe that the current extrapolation serves as a reliable estimate of the converged cold curve obtained from AFQMC for the current 4-atom cell studied in this work.

\subsection{Comparison to other methods}

\begin{figure}
  \includegraphics[width=0.48\textwidth]{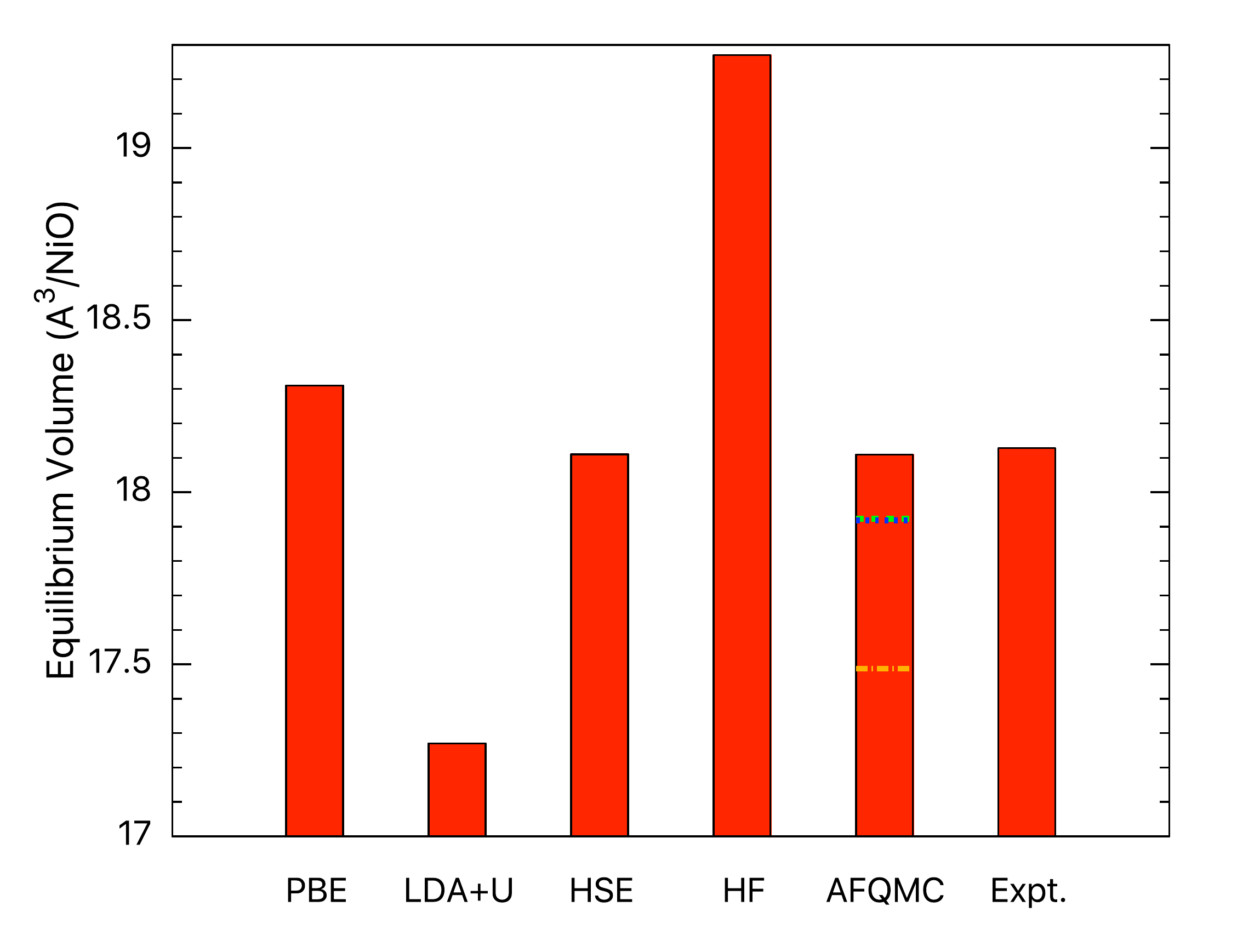}
  \includegraphics[width=0.48\textwidth]{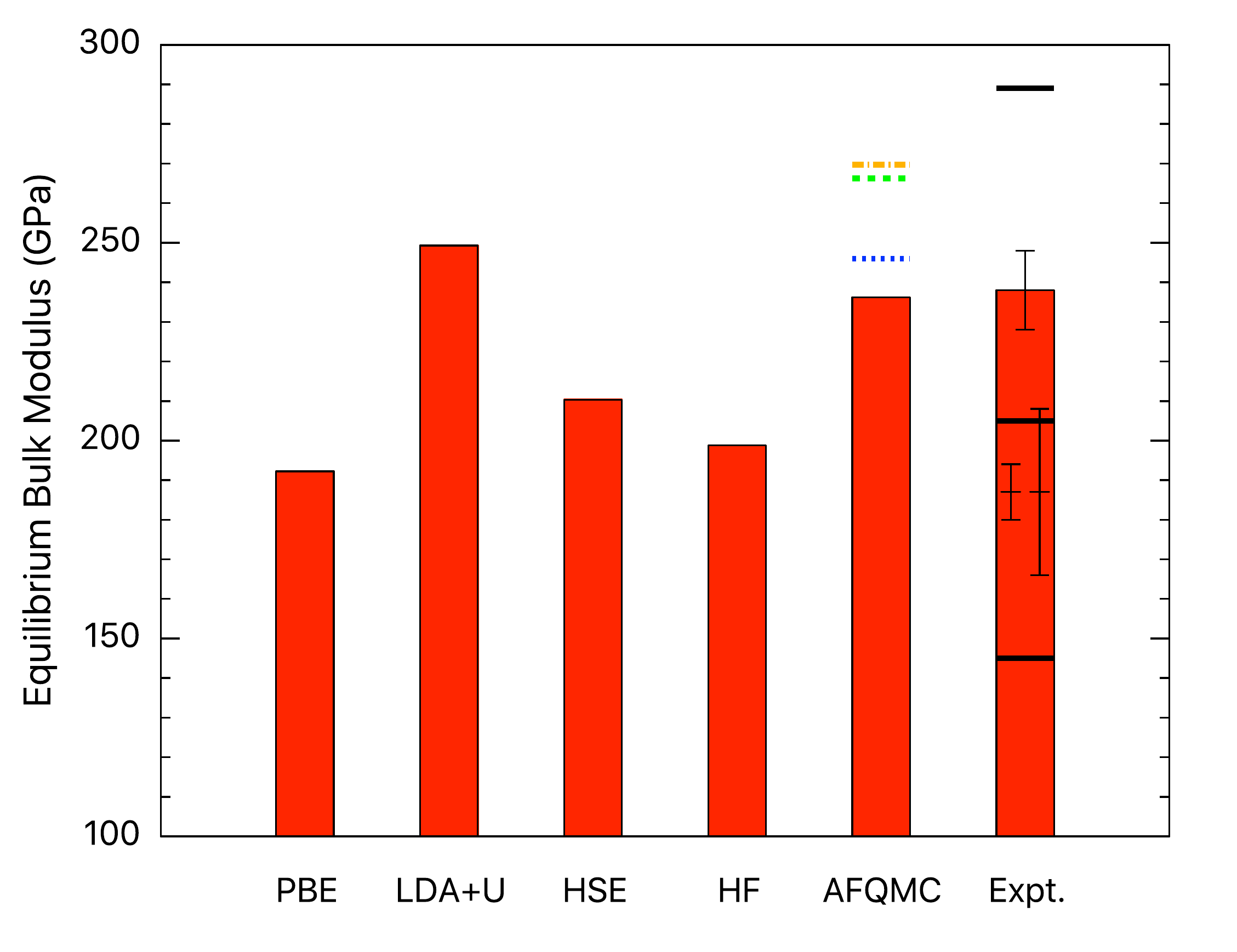}
  \caption{Comparison of the equilibrium volume and bulk modulus by fitting the cold curve from various methods to Murnaghan equation of state~\cite{Murnaghan244}. The scattered bars and data points in experimental bulk modulus denote different measurements~\cite{LDAU1998,Huang1994,Huang1995}. AFQMC values shown with the red rectangles are the CBS limits obtained by extrapolating the DZVP and TZVP values using \cref{eq:cbs}.
The yellow dash-dotted, green dashed, and blue dotted bars denote corresponding AFQMC values using the DZVP, TZVP, and TZV2P basis, respectively.}
  \label{fig:compareV0B0}
\end{figure}

\begin{table}[h]
\caption{\label{tab:eosNiO} AFQMC values (CBS limits obtained by extrapolating the DZVP and TZVP values using \cref{eq:cbs}) for the equilibrium volume $V_0$ and bulk modulus $B_0$ of NiO in rock-salt structure and anti-ferromagnetic state in comparison with experimental measurements and those from Hartree-Fock (HF) and DFT calculations, including  PBE, localized density approximation plus Hubbard $U$ (LDA+$U$), PBE+$U$, and Heyd-Scuseria-Ernzerhof (HSE06) hybrid functional. Diffusion Monte Carlo (DMC) and selected DFT or HF simulations from literature~\cite{krogelDMC} are also shown for comparison.}
    \centering
    \begin{ruledtabular}
    \begin{tabular}{lcc}
         & $V_0$ (\AA$^3$/NiO) & $B_0$ (GPa) \\
         \hline
      AFQMC    & 18.11 & 236 \\
      DMC~\cite{krogelDMC} & 17.96$\pm$0.04 & 196$\pm$4 \\
        PBE & 18.31 & 192 \\
        PBE~\cite{Tran2006} & 18.52 & 197 \\
        PBE~\cite{Feng2004} & 18.30 & 201 \\
        PBE~\cite{Feng2004} & 18.28 & 217 \\
        LDA~\cite{krogelDMC} & 16.73 & 232 \\
        LDA~\cite{Tran2006} & 16.85 & 257 \\
        PBE+U & 18.74 & 214 \\
        LDA+U & 17.27 & 249 \\
        LDA+U~\cite{krogelDMC} & 17.23 & 236 \\
        LDA+U~\cite{Tran2006} & 17.48 & 234 \\
        HSE06 & 18.11 & 210 \\
        HSE06~\cite{krogelDMC} & 17.98 & 198 \\
        PBE0~\cite{Tran2006} & 19.06 & 187 \\
        B3LYP~\cite{Feng2004} & 18.76 & 209 \\
        B3LYP~\cite{Feng2004} & 18.85 & 198 \\
        B3PW91~\cite{Tran2006} & 18.65 & 203 \\
        Fock-0.35~\cite{Tran2006} & 17.87 & 227 \\
        Fock-0.5~\cite{Tran2006} & 18.26 & 218 \\
        HF & 19.27 & 200 \\
        HF~\cite{LDAU1998} & 19.33 & -- \\
        Experiment    & 18.13\cite{Bartel1971} & 166-208~\cite{Huang1994},145,205,289~\cite{LDAU1998} \\
                      & (0 K, $a_0$=4.171 \AA) & 187$\pm$7~\cite{Huang1995},238$\pm$10~\cite{Huang1995} \\
    \end{tabular}
    \end{ruledtabular}
\end{table}

We obtained the NiO equilibrium volume ($V_0$) and bulk modulus ($B_0$) 
using a Murnaghan fit to the size-corrected AFQMC data\citep{Murnaghan244}, and used
 \cref{eq:cbs} to extrapolate the resulting energies to the CBS limits 
using the corresponding DZVP and TZV2P calculations.
The results are summarized in Table~\ref{tab:eosNiO} and
Fig.~\ref{fig:compareV0B0}.
We compare our results to UHF and spin-polarized DFT simulations for the same four-atom cell calculated using VASP\cite{vasp,paw,dft_sims}.
To investigate the importance of the exchange correlation functional we tested the PBE~\cite{PBE1996} and Heyd-Scuseria-Ernzerhof (HSE06)~\cite{HSE2006} functionals as well as the LDA+$U$~\cite{LDAca1980,LDApz1981,LDAU1998} and PBE+$U$ approaches.
Our DFT and UHF results agree well with those from previous publications~\cite{LDAU1998,Tran2006,Feng2004}.

We see from \cref{fig:compareV0B0} that, in the CBS limit, AFQMC provides remarkably consistent results for both the equilibrium volume and bulk modulus, despite the possible remaining errors due to the use of KZK  and basis set corrections.
In contrast, the PBE, LDA$+U$ and HF results give significantly varied results.
Overall, and as expected, DFT results exhibit a strong dependence on the choice of the exchange correlation functional.
Of the functionals tested, the HSE06 functional performs best when compared with both the DMC results of Ref.~\onlinecite{krogelDMC} and the experimental equilibrium volume.
The experimental data for the bulk modules is quite scattered so no real comparison can be made here.

\section{\label{sec:conclude}Conclusion}
In summary, we presented the application of the phaseless AFQMC method to a real, strongly correlated solid using periodic Gaussian basis sets.
We investigated the importance of size corrections on AFQMC energies and on structural properties.
We found that existing techniques to correct finite size errors in QMC work well even in strongly correlated materials and can be used in future studies on larger simulation cells.
We present a detailed analysis of the influence of basis set on the structural properties of NiO in the AFM II state, obtaining results that are reasonably converged
with respect to basis set size. We employ basis set extrapolation to obtain a correction for the energy missing when using our largest basis set, 
which we believe provides a meaningful estimate to the converged cold curve of NiO. We obtain excellent agreement with experimental measurements
on the equilibrium volume.
While these results are quite encouraging, this represent only the first step in a long journey whose final goal is the positioning of AFQMC as a method of choice 
in the study of strongly correlated materials. Ongoing work on NiO includes the study of larger basis sets and correlation-consistent effective-core potentials\citep{ncsuecp},
the use of larger unit cells to eliminate the need for size correction schemes, and the study of other properties including spin gaps, excitation energies and the interplay of magnetism and bang-gap closure.
Nonetheless, we believe that these preliminary calculations serve as a stepping stone in this direction.


{\it Acknowledgement.} We would like to thank Shiwei Zhang and Mario Motta for helpful discussions and Qiming Sun for assistance in running PySCF. S.Z. is in debt to Edgar Landinez for helpful discussions. 
This work was performed under the
auspices of the U.S. Department of Energy (DOE) by LLNL
under Contract No. DE-AC52-07NA27344.
Funding support was from the U.S. DOE, 
Office of Science, Basic Energy Sciences, 
Materials Sciences and Engineering Division, 
as part of the Computational Materials Sciences Program 
and Center for Predictive Simulation of Functional Materials (CPSFM).
Computer time was provided by the
Argonne Leadership Computing 
and Livermore Computing Facilities.


%
\end{document}